\documentclass[a4paper,twoside,twocolumn,aps,prl,superscriptaddress]{revtex4}
\usepackage{graphicx}
\usepackage{amsmath}
\usepackage{braket}
\usepackage{xcolor}

\begin{document}
\title{Measurements of the Correlation Function\\
of a Microwave Frequency Single Photon Source}
{\author{D. Bozyigit} \affiliation{Department of Physics, ETH Z\"urich, CH-8093, Z\"urich, Switzerland.}
\author{C.~Lang}
\affiliation{Department of Physics, ETH Z\"urich, CH-8093, Z\"urich, Switzerland.}
\author{L.~Steffen}
\affiliation{Department of Physics, ETH Z\"urich, CH-8093, Z\"urich, Switzerland.}
\author{J.~M.~Fink}
\affiliation{Department of Physics, ETH Z\"urich, CH-8093, Z\"urich, Switzerland.}
\author{M.~Baur}
\affiliation{Department of Physics, ETH Z\"urich, CH-8093, Z\"urich, Switzerland.}
\author{R.~Bianchetti}
\affiliation{Department of Physics, ETH Z\"urich, CH-8093, Z\"urich, Switzerland.}
\author{P.~J.~Leek}
\affiliation{Department of Physics, ETH Z\"urich, CH-8093, Z\"urich, Switzerland.}
\author{S.~Filipp}
\affiliation{Department of Physics, ETH Z\"urich, CH-8093, Z\"urich, Switzerland.}
\author{M.~P.~da~Silva}
\affiliation{D\'epartement de Physique, Universit\'e de Sherbrooke, Sherbrooke, Qu\'ebec, J1K 2R1 Canada.}
\author{A.~Blais}
\affiliation{D\'epartement de Physique, Universit\'e de Sherbrooke, Sherbrooke, Qu\'ebec, J1K 2R1 Canada.}
\author{A.~Wallraff}
\affiliation{Department of Physics, ETH Z\"urich, CH-8093, Z\"urich, Switzerland.}
\date{\today}
\begin{abstract}\bf
At optical frequencies the radiation produced by a source, such as a laser, a black body or a single photon source, is frequently characterized by analyzing the temporal correlations of emitted photons using single photon counters. At microwave frequencies, however, there are no efficient single photon counters yet. Instead, well developed linear amplifiers allow for efficient measurement of the amplitude of an electromagnetic field. Here, we demonstrate how the properties of a microwave single photon source can be characterized using correlation measurements of the emitted radiation with such detectors. We also demonstrate the cooling of a thermal field stored in a cavity, an effect which we detect using a cross-correlation measurement of the radiation emitted at the two ends of the cavity.
\end{abstract}
\maketitle

In quantum optics \cite{Walls1994}, the single photon detector is a versatile tool to explore the properties of radiation emitted from a variety of classical and quantum sources. At optical frequencies these detectors easily produce a `click' when a single photon impinges on them. Recording the statistics of such events, one is able to measure not only the average of the number of emitted photons $n = \langle a^{\dagger} a\rangle$ but also higher order statistical correlations between the emitted photons. Here, $a^{\dagger}$ and $a$  are the creation and annihilation operators of the radiation detected in a given mode. At lower frequencies, such as in the microwave frequency domain, however, no efficient single photon detectors exist to date
as photons at these frequencies carry an energy that is orders of magnitude less than at optical frequencies. Instead, at microwave frequencies linear amplifiers followed by an instrument recording a voltage, such as an oscilloscope, are commonly used to detect small amplitude electromagnetic fields. Such amplifiers have several orders of magnitude gain but also add noise to the signal. Nevertheless, valuable information about the properties of quantum radiation sources in the microwave domain can be acquired. With sufficient averaging, electric field amplitudes, which are proportional to the sum and difference of the field operators $a$ and $a^{\dagger}$, can be detected in this way even on the single photon level \cite{Houck2007}. We demonstrate here that by recording the time series of the detected signal -- instead of its time average -- and using efficient digital signal processing, we can extract characteristic correlation functions of radiation generated by microwave frequency emitters such as a single photon source or a black body.}

For our experiments we have realized a microwave frequency single photon source in a superconducting electronic circuit, similar to the one presented in \cite{Houck2007}. In our circuit, we coherently and controllably couple a single qubit to a high quality resonator to create an individual photon on demand. The superconducting transmon qubit \cite{Koch2007} used in this experiment is characterized by its maximum Josephson energy $E_{\rm{J,max}} \approx 14.4 \, \rm{GHz}$, its charging energy $E_{\rm{C}} \approx 500 \, \rm{MHz}$ and its energy relaxation and dephasing times in excess of a few hundred nanoseconds. The transition frequency of the qubit is flux tunable using both a quasi-static magnetic field generated with a miniature coil and an on chip transmission line to generate nanosecond time scale flux pulses. By integrating our qubit into a superconducting coplanar transmission line resonator of frequency $\nu_{\rm{r}} = 6.433 \, \rm{GHz}$ and quality factor $Q = 2060$ we couple it strongly to a single mode $a$ of the radiation field stored in the resonator. This approach is known as circuit quantum electrodynamics~\cite{Wallraff2004b} and allows to study in exquisite detail the interaction of quantum two-level systems with quantized radiation fields.

\begin{figure}[!t]
\centering
\includegraphics[width=1 \columnwidth]{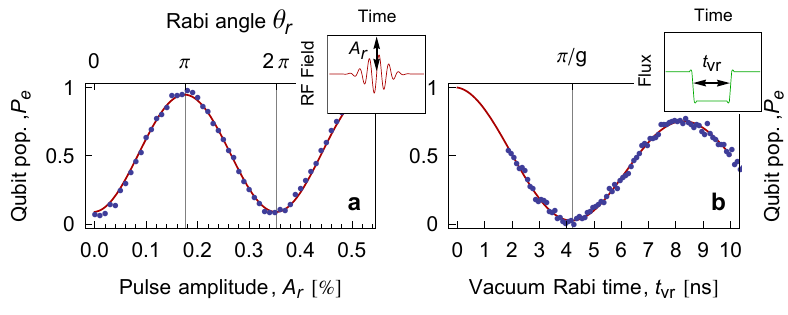}%
\caption{\textbf{Control of qubit state and qubit/resonator interaction.} \textbf{a}, Qubit excited state population $P_{\rm{e}}$ versus Rabi pulse amplitude $A_{\rm{r}}$. The Rabi rotation angle $\theta_{\rm{r}}$ is indicated on the top axis. The used pulse shape  is shown in the inset. \textbf{b}, $P_{\rm{e}}$ versus qubit-resonator interaction time $t_{\rm{vr}}$ with the used flux pulse shape in inset. Note that the interaction time can experimentally not be made shorter than $1.9 \, \rm{ns}$ due to finite rise time of the pulses. Dots are data, lines fits to theory.} \label{fig:rabi}
\end{figure}

Applying a phase controlled truncated gaussian microwave pulse of variable amplitude $A_{\rm{r}}$ and total duration $t_{\rm{r}} = 18 \, \rm{ns}$ to the qubit biased at a transition frequency of $\nu_{\rm{q}} = 6.933 \, \rm{GHz}$, we prepare a superposition state $|\psi_{\rm{q}}\rangle = \alpha |g\rangle + \beta |e\rangle$ between the qubit ground $|g\rangle$ and excited states $|e\rangle$. The superposition is characterized by the two complex probability amplitudes $\alpha =  \cos(\theta_{\rm{r}}/2)$ and $\beta =  \sin(\theta_{\rm{r}}/2)e^{i\phi}$ which are parameterized by the polar (Rabi) angle $\theta_{\rm{r}}$ and the phase angle $\phi$. We characterize the prepared qubit state using a pulsed dispersive measurement of the resonator transmission \cite{Bianchetti2009} and clearly observe Rabi oscillations in the qubit population $P_{\rm{e}}$ versus the amplitude $A_{\rm{r}}$ (Fig.~\ref{fig:rabi}a). After the qubit state preparation, we apply a current pulse of controlled amplitude and duration to the flux bias line to tune the qubit transition frequency into resonance with the resonator frequency $\nu_{\rm{r}}$. We time-resolve the resonant vacuum Rabi oscillations of the coupled system at a frequency of $g / \pi = 118 \, \rm{MHz}$ by dispersively measuring the qubit state after it has been tuned back to the frequency $\nu_{\rm{q}}$ strongly detuned from the resonator (Fig.{\ref{fig:rabi}}b). Adjusting the qubit-resonator interaction time to half a vacuum Rabi period $t_{\rm{vr}} = \pi / g = 4.2 \, \rm{ns} $, we coherently map the qubit state $|\psi_{\rm{q}}\rangle$ to an equivalent superposition state $|\psi_{\rm{c}}\rangle = \alpha |0\rangle + \beta |1\rangle$ of the $|0\rangle$ and $|1\rangle$ photon Fock states stored in the resonator mode $a$. Similar techniques have been used to prepare and measure a wide range of intra-cavity photon superposition states in recent experiments both with superconducting circuits \cite{Hofheinz2009} and with Rydberg atoms \cite{Deleglise2008}.

\begin{figure}[t!]
\centering
\includegraphics[width=0.7 \columnwidth]{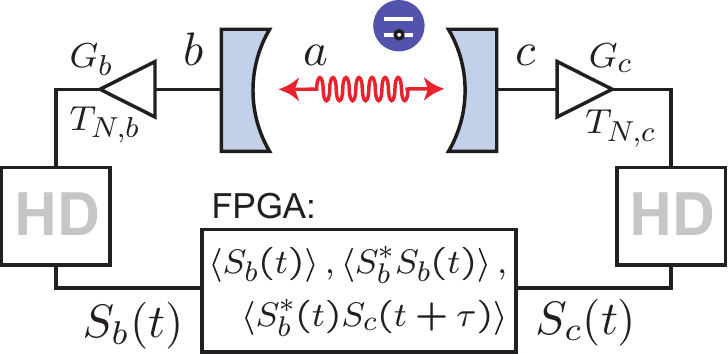}%
\caption{\textbf{Schematic of the experimental setup.} Resonator in which a two level system emits a photon into the mode $a$. Linear amplifiers with gain $G_{b,c}$ couple symmetrically to the cavity outputs amplifying the radiation in output modes $b$ and $c$. Homodyne detectors (HD) 
extract both quadratures of the output fields and feed results to field programmable gate array (FPGA) based digital correlation electronics.} \label{fig:ExperimentSchematics}
\end{figure}

\begin{figure}[t!]
\centering
\includegraphics[width=1 \columnwidth]{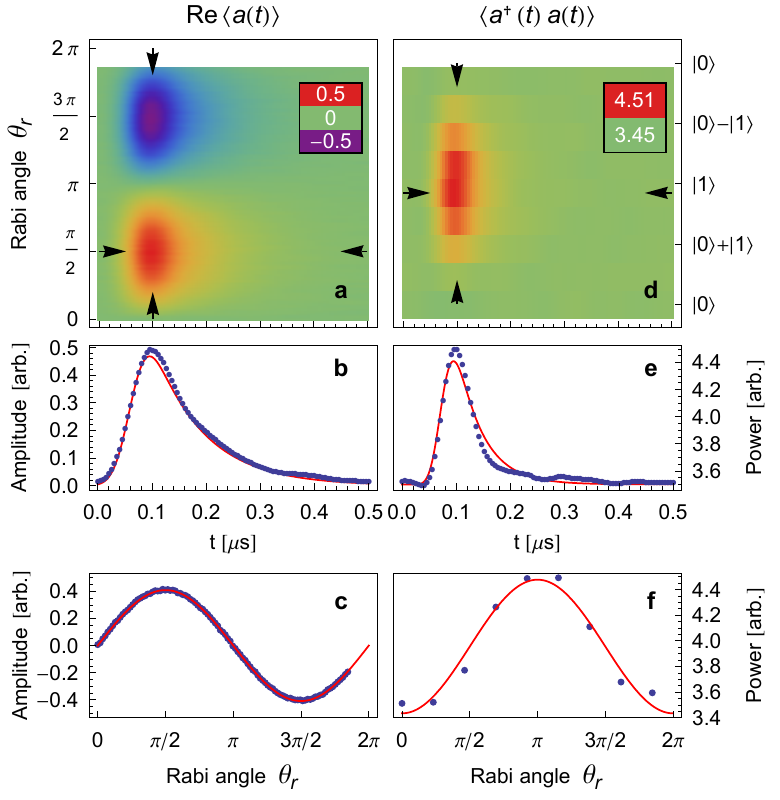}
\caption{\textbf{Quadrature amplitude and cross-power measurements.} \textbf{a}, Measured time dependence of cavity field quadrature amplitude of mode $b$ for zero and one photon superposition states characterized by the Rabi angle $\theta_{\rm{r}}$ (left axis) and generated state (right axis). \textbf{b}, Single quadrature trace at $\theta_{\rm{r}}=\pi/2$ in \textbf{a} corresponding to $(\ket{0}+\ket{1})/\sqrt{2}$. \textbf{c}, Dependence of maximum quadrature amplitude on $\theta_r$. \textbf{d}, Measured time dependence of cross-power between modes $b$ and $c$ for the same preparation as in \textbf{a}. \textbf{e}, Single power trace at $\theta_{\rm{r}}=\pi$ in \textbf{c} corresponding to $\ket{1}$. \textbf{f}, Dependence of maximum cross-power on $\theta_r$. Blue dots are data, red line is theory, see text.} \label{fig:QuadPowerPlots}
\end{figure}

In a next step, we characterize zero and one photon superposition states by measuring the quadrature amplitudes of the microwave fields emitted from \emph{both} ends of the cavity, see schematic in Fig.~\ref{fig:ExperimentSchematics}. For this purpose we have realized a symmetric circuit QED setup in which the output fields are detected independently and simultaneously at both ports of the cavity. It is interesting to note that this setup is essentially equivalent to one in which the output of a radiation source is investigated at the two output ports of a beam splitter \cite{Silva2010}.
Our scheme comprises two independent detection chains similar to the one pioneered by Gabelli \emph{et al.} \cite{Gabelli2004}. 
Each chain consists of a cold amplifier with gain $G_{b,c} \approx 33 \, \rm{dB}$ and noise temperature $T_{\rm{N}(b,c)} \approx 4.5 \, \rm{K}$ followed by a two stage heterodyne detector in which the signal is down converted from the resonator frequency to $25 \, \rm{MHz}$ in an analog stage and to d.c.~in a digital homodyne stage. This allows for the measurement of the electric field at both ends of the resonator characterizing the radiation emitted into the modes labeled $b$ and $c$. In this way, we extract the complex envelope $S_{b,c}(t)$ of the amplified electric field $E^{(+)}_{b,c}$ described by
\begin{align}
E^{(+)}_{b,c} (t) + N_{b,c}(t) = S_{b,c}(t) e^{-2 \pi i \nu_{\rm{r}} t}
\end{align}
where the real and imaginary part of $S_{b,c}(t)$ are the two field quadratures in the frame rotating at the resonator frequency and $N_{b,c}(t)$ is the noise added by the amplifier. Using input-output theory \cite{Gardiner1985}, one can show that the full information about the intra-cavity mode $a$ can be extracted from a measurement of the propagating modes $b$ and $c$ \cite{Silva2010}.
As a first example, we present a measurement of the time dependence of both quadrature amplitudes of the electric field of the output mode $b$ at one end of the resonator. This gives us access to the expectation value of the annihilation operator of the cavity field $\langle S_{b}(t) \rangle \propto \langle a(t) \rangle$ \cite{Silva2010}. Similar measurements were presented in Ref.~\cite{Houck2007}, where the cavity photon was created by Purcell limited spontaneous emission. Figure \ref{fig:QuadPowerPlots}a shows the real part of $\langle S_{b}(t) \rangle$ versus time $t$ after the preparation of the photon superposition state $|\psi_{\rm{c}}\rangle$ characterized by the qubit Rabi angle $\theta_{r}$ used for its preparation.
We find excellent agreement with the expected average field quadrature amplitude $\langle a\rangle \propto \sin{(\theta_r)}/2$ (Fig.~\ref{fig:QuadPowerPlots}c), where we find the largest signals for the superposition states $|\psi^+_{\rm{c}}\rangle = (|0\rangle+|1\rangle)/\sqrt{2}$ and $|\psi^-_{\rm{c}}\rangle = (|0\rangle-|1\rangle)/\sqrt{2}$ prepared using $\theta_{r} = \pi/2$ and $3\pi/2$, respectively. As expected from the uncertainty principle, the Fock states $|0\rangle$ and $|1\rangle$ prepared with $\theta_{r} = 0$ and $\pi$, respectively, do not show any quadrature amplitude signals (Fig.~\ref{fig:QuadPowerPlots}a) since their phase is completely uncertain. For all of the above measurements, the overall global phase of the signals is adjusted such that the imaginary part of $S_{b}(t)$ is equal to zero which therefore is not displayed. We also note that the amplifier noise averages to zero in the quadrature amplitude measurement. Moreover, the time dependence of all measurement traces is well understood. For the state $|\psi^+_{\rm{c}}\rangle$ (Fig.~\ref{fig:QuadPowerPlots}b), for example, the characteristic decay time is given by twice the cavity decay time $2 T_{\kappa} = Q / \pi \nu_r = 102 \, \rm{ns}$. The rise time, which should ideally be $t_{\rm{vr}} = 4.2 \, \rm{ns}$, is limited by the bandwidth $14 \, \rm{MHz}$ of our detection scheme.

In our measurement scheme, we simultaneously record the time dependent quadrature amplitudes $S_{b,c}(t)$ detected at both ports of the cavity for \emph{each} single photon that we generate. This is realized using a two channel analog-to-digital converter (ADC) with a time resolution of $10 \, \rm{ns}$. Based on the measurement record of each event, we can then calculate \emph{any} expectation value, such as averages, products or correlations, that can be expressed in terms of the detected output signals $S_{b,c}(t)$. Processing this data in real time using field programmable gate array (FPGA) electronics allows us to efficiently extract information even in the presence of substantial noise added by the amplifier.

As a first experiment taking advantage of this scheme, we have measured the expectation value of the instantaneous power $\langle S_b^*(t) S_b(t)\rangle$ emitted into the output mode $b$ with the cavity mode $a$ prepared in the Fock state $|1\rangle$ (not shown). It is important to note that we have digitally calculated the power by multiplying the quadrature amplitudes $S_b(t)$ for each single photon pulse instead of using a diode as a power meter in which the detection and the averaging is realized within the detector \cite{Houck2007}. As we will see in the following, this digital procedure is very versatile as it also allows to calculate cross-powers and cross-correlations digitally. In the direct power measurement, the detected noise power of the amplifier dominates by a factor of about $2000$ over the single photon power which is still observed using sufficient averaging (similar as in Fig.~\ref{fig:QuadPowerPlots}e).
From the background noise we determine the system noise temperature $T_{\rm{N(sys)}} \approx 17 \, \rm{K}$ of our detection chain with respect to the output of the resonator. $T_{\rm{N(sys)}}$ is substantially higher than the noise temperature of the amplifiers because of absorption in the cables and insertion loss of components in the detection chain.

Calculating the cross-power $\langle S_b^*(t)S_c(t)\rangle$ between the two output modes instead of the direct power emitted into just a single output mode, we can reject most of the noise added by the amplifiers, demonstrating the versatility of our digital scheme. The detected cross-power is related to the average photon number in the cavity as \cite{Silva2010}
\begin{align}
\langle S^*_b(t) S_c(t) \rangle \propto \langle a^\dagger(t) a (t) \rangle + P(N_{bc}),
\end{align}
where $P(N_{bc})$ is the power of correlated noise between channels $b$ and $c$. 
In these measurements, the detected noise cross-power is a factor of $1/500$ smaller than the direct noise power of each amplifier as the two detection chains add predominantly uncorrelated noise. The residual correlations are of technical origin, such as insufficient isolation of the two amplifier chains, correlated digitizer noise and residual resonator thermal noise due to incomplete thermalization of the resonator inputs. We have characterized the measured cross-power of our single photon source for the same set of cavity superposition states as used for the quadrature amplitude measurements (Fig.~\ref{fig:QuadPowerPlots}c). We find excellent agreement of the temporal evolution of the cavity photon number in dependence on the preparation angle of the photon state $\langle a^\dagger a \rangle \propto (1-\cos\theta_r)/2$ (Fig.~\ref{fig:QuadPowerPlots}f). The maximum cross-power is measured for the Fock state $|1\rangle$ ($\theta_r = \pi$) and the minimum power for the $|0\rangle$ state ($\theta_r = 0$ or $2\pi$) (Fig.~\ref{fig:QuadPowerPlots}d).

\begin{figure}[t!]
\centering
\includegraphics[width=1 \columnwidth]{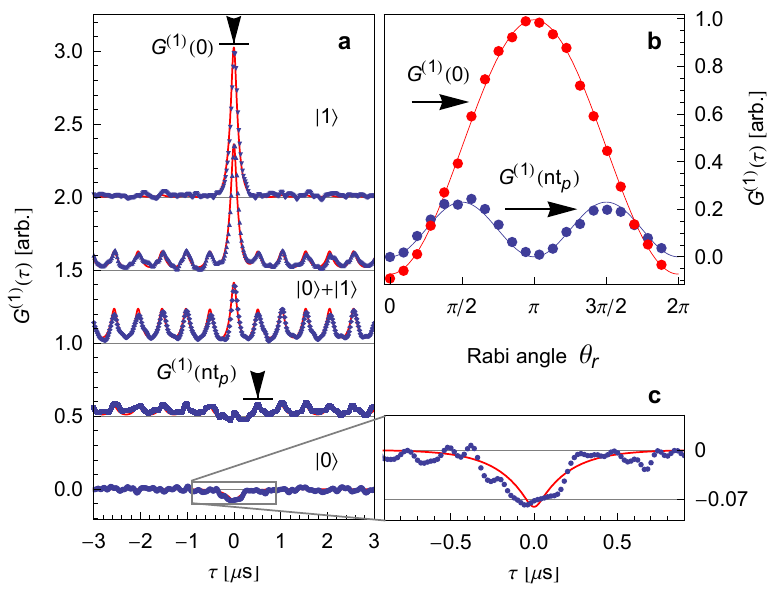}%
\caption{\textbf{Correlation function measurements.} \textbf{a}, Time dependence of first-order correlation function of the cavity field $G^{(1)}$ for indicated states. Data is offset for improved visibility. \textbf{b}, Correlation function at $\tau = 0$ and $\tau = n t_p$ versus $\theta_{\rm{r}}$. Dots are data, lines are theoretical predictions. \textbf{c}, Enlarged trace for $\theta_{\rm{r}}=0$ from \textbf{a} displaying cavity field cooling.} \label{fig:G1Plots}
\end{figure}

Finally, we have characterized our single photon source using time-dependent first-order cross-correlation measurements of the two output modes of the resonator
\begin{align}
\Gamma^{(1)}(\tau) = \int \langle S^{*}_b(t) S_c(t+\tau) \rangle dt \, .
\end{align}
For this purpose, we generate a train of $40$ single photon pulses, each created using the procedure described above, with a pulse separation of $t_{\rm{p}} = 512 \, \rm {ns}$ which is much greater than the qubit and cavity decay times. To remove the background, we subtract the measured correlation function $\Gamma^{(1)}_{ss}(\tau)$ in the resonator steady-state from the signal acquired when performing the photon state preparation sequence $\Gamma^{(1)}(\tau)$. From the recorded quadrature amplitude data, we calculate in real-time
\begin{align}
\Gamma^{(1)}(\tau) - \Gamma^{(1)}_{ss}(\tau) \propto G^{(1)}(\tau)\, ,
\end{align}
which gives us access to the first-order correlation function $G^{(1)}(\tau) = \int \langle a^\dagger (t) a(t+\tau) \rangle dt$ of the resonator field~\cite{Silva2010}. To measure each trace in Fig.~\ref{fig:G1Plots}a,
$128 \times 10^6$ trains of 40 photons were prepared in a specific state and $G^{(1)}(\tau)$ was calculated in real time using our FPGA based electronics, corresponding to more than 1 terabyte of data that have been evaluated in approximately $1$ hour.

The correlation function data $G^{(1)}(\tau)$ (Fig.\ref{fig:G1Plots}a) is characterized by a set of peaks that are separated by the repetition time $t_{\rm{p}}$ of the single photon source. The amplitude of $G^{(1)}$ at $\tau = 0$ and $\tau = n t_{\rm{p}}$, representing the correlation between a pulse $i$ and $i+n$, depends in a characteristic fashion on $\theta_{\rm{r}}$. For the Fock state $|1\rangle$ (at $\theta_{r} = \pi$), the correlation function $G^{(1)}(0)$ is at a maximum and vanishes at $G^{(1)}(n t_{\rm{p}})$ as there is no coherence between photons emitted from the source at different times. In fact, $G^{(1)}(0) \propto \langle a^{\dagger} a\rangle \propto |\beta|^2 = \sin(\theta_r/2)^2$ oscillates sinusoidally with the preparation angle, as it essentially measures the average photon number of the generated field (Fig.~\ref{fig:G1Plots}b). For photon superposition states, the expectation values of $\langle a^{\dagger}\rangle$ and $\langle a\rangle$ of subsequently generated photon states have non-vanishing values, as discussed before. Since photons from different repetitions of the experiments are uncorrelated, $G^{(1)}(n t_{\rm{p}})\propto \langle a^{\dagger}\rangle\langle a\rangle$ which has a finite value and oscillates at half the period. Thus, $G^{(1)}(n t_{\rm{p}})\propto |\alpha\beta|^2 = \sin(\theta_r)^2/4$ is maximized for the states $|\psi^{+}_{\rm{c}}\rangle$ and $|\psi^{-}_{\rm{c}}\rangle$ (Fig.~\ref{fig:G1Plots}b).

The observed features of the first-order correlation function
are characteristic for the non-classical radiation generated by our
deterministic single photon source. Considering the well controlled procedure implemented for generating the single photon pulses any other pulsed coherent or thermal field can essentially be ruled out as being the source of the measured correlations.
This indicates for our scheme that the quantum mechanical properties of the radiation are not lost in the detection chain but are fully retained in the form of statistical information of the data and can be extracted by appropriate single-shot analysis. Measurements of the second order correlation function $G^{(2)}$ would provide unambiguous proof of the quantum character of the field, independent of any prior knowledge about its source \cite{Walls1994}. So far, the long integrations times required due to the presence of noise added by the amplifiers have prevented their successful measurement.

Interestingly, our time-dependent first-order correlation function measurements allow us to observe cooling of the resonator field through its interaction with the qubit in its ground state. 
If the thermal occupation of the qubit excited state is substantially smaller than the thermal occupation of the resonator, the qubit can absorb a photon from the resonator during the interaction time $t_{\rm{vr}}$. The qubit then emits the photon into the environment at a different frequency thereby cooling the resonator field.
The cooling is observable as a pronounced dip in the measured $G^{(1)}(\tau)$ around $\tau = 0$, see Fig.~\ref{fig:G1Plots}c.
An analysis of the size of the dip in the correlation function allows us to extract the thermal background population of the resonator $n_{\rm{bg}} = 0.07$ corresponding to a field temperature of $T_{\rm{bg}} = 115 \, \rm{mK}$ which is in agreement with independent measurements of $T_{\rm{bg}}$ from the vacuum Rabi mode splitting spectrum \cite{Fink2009b}.
Analyzing in detail the time dependent correlation function measurements of the single photon source in the presence of a weak thermal background field, we find excellent agreement between our data and theory \cite{Silva2010} for all prepared field states (see solid lines in Fig.~\ref{fig:G1Plots}).

Our experiments clearly demonstrate that correlation function measurements based on quadrature amplitude measurements are a powerful tool to characterize quantum properties of propagating microwave frequency radiation fields. Even in the presence of noise added by the amplifier, efficient data processing techniques allow for the measurements of higher statistical moments of the fields. When better, possibly quantum limited, amplifiers \cite{Castellanos-Beltran2008} become available the demonstrated techniques may help to enable the full tomography of propagating radiation fields. Furthermore, the flexibility of circuit design and the high level of control achievable in circuit QED will enable a variety of future experiments with quantum microwave fields for basic research and applications.

While preparing this manuscript, we became aware of related work by Menzel \emph{et al.} \cite{Menzel2010}.

\begin{acknowledgments}
We thank  T.~Frey and G.~Littich for their contributions at the early stages of the project. This work was supported by ERC, and ETHZ. M.P.S. was supported by a NSERC postdoctoral fellowship. A.B. was supported by NSERC, CIFAR, and the Alfred P. Sloan Foundation.
\end{acknowledgments}



\bibliographystyle{nature}
\bibliography{./QudevRefDB}

\begin{thebibliography}{10}

\bibitem{Walls1994}
Walls, D. and Milburn, G.
\newblock {\em Quantum optics}.
\newblock Spinger-Verlag, Berlin,  (1994).

\bibitem{Houck2007}
Houck, A., Schuster, D., Gambetta, J., Schreier, J., Johnson, B., Chow, J.,
  Frunzio, L., Majer, J., Devoret, M., Girvin, S., and Schoelkopf, R.
\newblock {\em Nature}{ \bf 449}, 328 (2007).

\bibitem{Koch2007}
Koch, J., Yu, T.~M., Gambetta, J., Houck, A.~A., Schuster, D.~I., Majer, J.,
  Blais, A., Devoret, M.~H., Girvin, S.~M., and Schoelkopf, R.~J.
\newblock {\em Physical Review A}{ \bf 76}(4), 042319 (2007).

\bibitem{Wallraff2004b}
Wallraff, A., Schuster, D.~I., Blais, A., Frunzio, L., Huang, R.~S., Majer, J.,
  Kumar, S., Girvin, S.~M., and Schoelkopf, R.~J.
\newblock {\em Nature}{ \bf 431}, 162--167 (2004).

\bibitem{Bianchetti2009}
Bianchetti, R., Filipp, S., Baur, M., Fink, J.~M., G\"{o}ppl, M., Leek, P.~J.,
  Steffen, L., Blais, A., and Wallraff, A.
\newblock {\em Physical Review A (Atomic, Molecular, and Optical Physics)}{ \bf
  80}(4), 043840 (2009).

\bibitem{Hofheinz2009}
Hofheinz, M., Wang, H., Ansmann, M., Bialczak, R.~C., Lucero, E., Neeley, M.,
  O'Connell, A.~D., Sank, D., Wenner, J., Martinis, J.~M., and Cleland, A.~N.
\newblock {\em Nature}{ \bf 459}(7246), 546--549 May  (2009).

\bibitem{Deleglise2008}
Deleglise, S., Dotsenko, I., Sayrin, C., Bernu, J., Brune, M., Raimond, J.-M.,
  and Haroche, S.
\newblock {\em Nature}{ \bf 455}(7212), 510--514 September  (2008).

\bibitem{Silva2010}
da~Silva, M.~P., Bozyigit, D., Wallraff, A., and Blais, A.
\newblock {\em (unpublished)}{ \bf } (2010).

\bibitem{Gabelli2004}
Gabelli, J., Reydellet, L.-H., Feve, G., Berroir, J.-M., Placais, B., Roche,
  P., and Glattli, D.~C.
\newblock {\em Physical Review Letters}{ \bf 93}(5), 056801 (2004).

\bibitem{Gardiner1985}
Gardiner, C.~W. and Collett, M.~J.
\newblock {\em Phys. Rev. A}{ \bf 31}(6), 3761-- June  (1985).

\bibitem{Fink2009b}
Fink, J.~M., Baur, M., Bianchetti, R., Filipp, S., G{\"o}ppl, M., Leek, P.~J.,
  Steffen, L., Blais, A., and Wallraff, A.
\newblock {\em Phys. Scr.}{ \bf T137}, 014013 (2009).

\bibitem{Castellanos-Beltran2008}
Castellanos-Beltran, M.~A., Irwin, K.~D., Hilton, G.~C., Vale, L.~R., and
  Lehnert, K.~W.
\newblock {\em Nature Physics}{ \bf 4}, 929--931 (2008).

\bibitem{Menzel2010}
Menzel, E.~P., Deppe, F., Mariantoni, M., Caballero, M.~A.~A., Baust, A.,
  Niemczyk, T., Hoffmann, E., Marx, A., Solano, E., and Gross, R.
\newblock arXiv:1001.3669,  (2010).

\end{thebibliography}

\end{document}